\documentclass{ws-procs975x65}

\newcommand{\al}{\ensuremath{\alpha} }
\newcommand{\be}{\ensuremath{\beta} }
\newcommand{\ga}{\ensuremath{\gamma} }
\newcommand{\la}{\ensuremath{\lambda} }
\newcommand{\Si}{\ensuremath{\Sigma} }
\newcommand{\psibar}{\ensuremath{\overline\psi} }
\newcommand{\vev}[1]{\ensuremath{\left\langle #1 \right\rangle} }
\newcommand{\pbp}{\ensuremath{\vev{\psibar\psi}} }
\newcommand{\gsim}{\ensuremath{\gtrsim} }
\newcommand{\lsim}{\ensuremath{\lesssim} }
\newcommand{\X}{\ensuremath{\!\times\!} }
\usepackage{cancel}
\newcommand{\Sb}{\ensuremath{\cancel{S^4}} }
\newcommand{\fig}[1]{Fig.~\ref{#1}}
\newcommand{\secref}[1]{Section~\ref{#1}}

\begin{document}
\title{Reaching the chiral limit in many flavor systems}
\author{Anna~Hasenfratz,$^*$ Anqi~Cheng, Gregory~Petropoulos and David~Schaich}
\address{Department of Physics, University of Colorado, Boulder, CO 80309, United States \\ $^*$E-mail: anna@eotvos.colorado.edu}

\begin{abstract} 
  We present a brief overview of our recent lattice studies of SU(3) gauge theory with $N_f = 8$ and 12 fundamental fermions, including some new and yet-unpublished results.
  To explore relatively unfamiliar systems beyond lattice QCD, we carry out a wide variety of investigations with the goal of synthesizing the results to better understand the non-perturbative dynamics of these systems.
  All our findings are consistent with conformal infrared dynamics in the 12-flavor system, but with 8 flavors we observe puzzling behavior that requires further investigation.

  Our new Monte Carlo renormalization group technique exploits the Wilson flow to obtain more direct predictions of a 12-flavor IR fixed point.
  Studies of $N_f = 12$ bulk and finite-temperature transitions also indicate IR conformality, while our current results for the 8-flavor phase diagram do not yet provide clear signs of spontaneous chiral symmetry breaking.
  From the Dirac eigenvalue spectrum we extract the mass anomalous dimension $\ga_m$, and predict $\ga_m^{\star} = 0.32(3)$ at the 12-flavor fixed point.
  The $N_f = 8$ system again shows interesting behavior, with a large anomalous dimension across a wide range of energy scales.
  We use the eigenvalue density to predict the chiral condensate, and compare this approach with direct and partially-quenched \pbp measurements.
\end{abstract}

\bodymatter
\section{Introduction and overview} 
The Higgs boson discovered at the Large Hadron Collider in 2012 remains consistent with the minimal standard model.\cite{CMS:2013lba}
Models of new strong dynamics in which the Higgs is a composite must describe this light scalar in order to remain phenomenologically viable.
Such strongly-coupled systems require non-perturbative analysis, which has inspired several recent large-scale lattice studies.
Lattice investigations beyond QCD not only need to identify promising models to consider, we also need to determine the best approaches to use; in many cases the most effective methods differ from the familiar techniques of lattice QCD.
Our group has recently focused on two models, $N_f = 8$ and 12 fundamental fermions interacting with SU(3) gauge fields.\cite{Hasenfratz:2011xn, Cheng:2011ic, Jin:2012dw, Lin:2012iw, Aoki:2012eq, Hasenfratz:2012fp, Schaich:2012fr, Deuzeman:2012ee, Fodor:2012uw, Fodor:2012et, Petropoulos:2012mg, Itou:2012qn, Cheng:2013eu, Aoki:2013pca, Aoki:2013xza, Giedt:2012LAT}
We have developed several improved methods to explore these and other models, focusing on systems likely to exhibit conformal or approximately-conformal dynamics in the infrared.

In these proceedings we briefly present some methods we developed and results we obtained for the 8- and 12-flavor models, including some new and preliminary findings that will be published in the future.
All our 12-flavor results are consistent with IR conformality, while with $N_f = 8$ we observe some puzzling behavior that we continue to investigate.
After reviewing the lattice phase diagrams for these models in \secref{sec:phases}, we present our new Wilson-flowed Monte Carlo renormalization group (MCRG) method in \secref{sec:WMCRG}.
In \secref{sec:eig} we summarize our recent studies of the Dirac eigenvalue spectrum, finally considering the chiral condensate in \secref{sec:pbp}.

Our studies use a plaquette gauge action that includes an adjoint term with coefficient $\be_A = -0.25\be_F$, and nHYP-smeared staggered fermions with smearing parameters $\al = (0.5, 0.5, 0.4)$.
In earlier work\cite{Cheng:2011ic, Schaich:2012fr} we showed that both the 8- and 12-flavor systems possess an unusual phase where the single-site shift symmetry of the staggered action is spontaneously broken (``$\Sb$'').
This strong-coupling lattice phase is quite generic and has been observed by all $N_f = 12$ investigations using improved staggered actions.\cite{Deuzeman:2012ee, Fodor:2012et}
First-order transitions separate the \Sb phase from the weak-coupling phase where the continuum limit is defined.
This prevents simulations from investigating stronger couplings.
For $N_f = 12$ we can still access a large coupling range, but in the 8-flavor case the presence of the \Sb phase prevents, at least for now, resolution of the puzzles we encounter in our studies.
In the following we present results up to the \Sb phase, but our focus is on the weak-coupling side.

\section{\label{sec:phases}The finite temperature phase structure} 
\begin{figure}[tb]
  \includegraphics[width=0.46\linewidth]{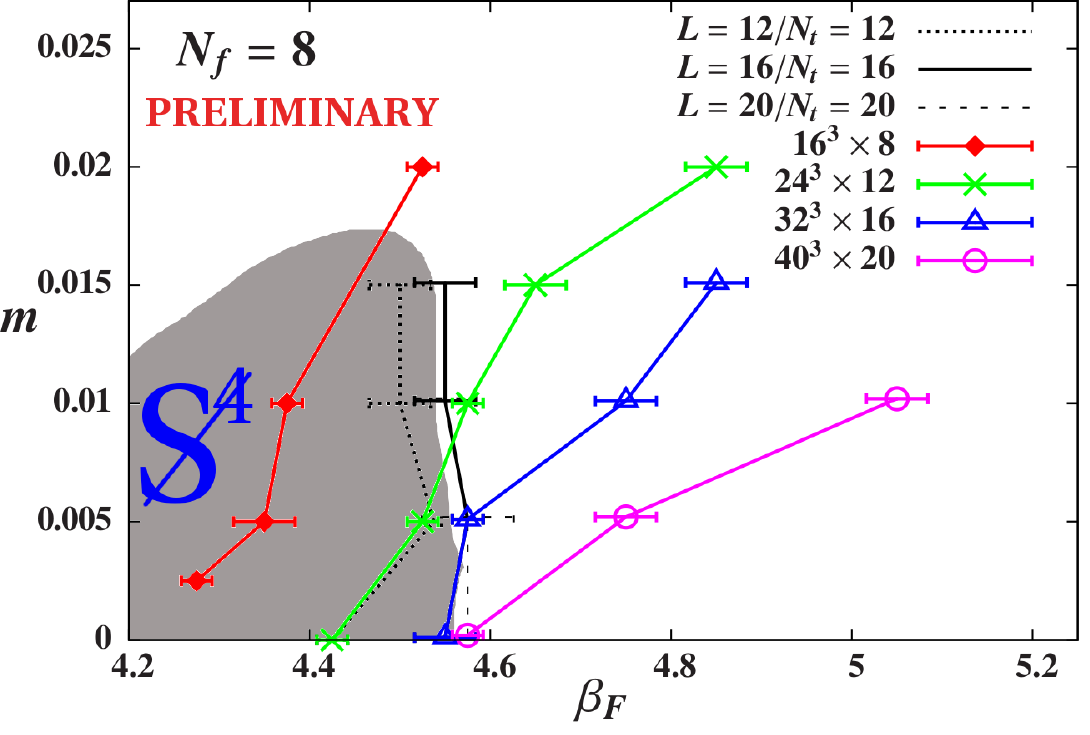}\hfill
  \includegraphics[width=0.46\linewidth]{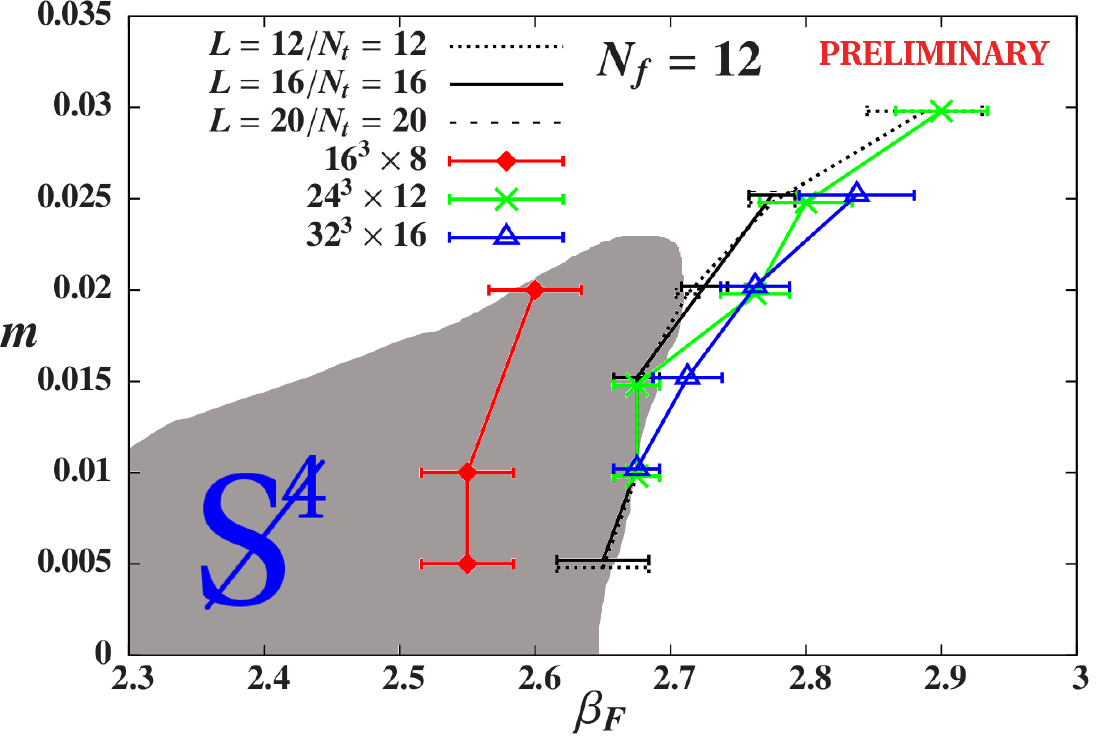}
  \caption{\label{fig:phases} Bulk and finite-temperature transitions for $N_f = 8$ (left) and $N_f = 12$ (right).\protect{\cite{Schaich:2012fr}}  The \Sb phase is shaded and the \Sb bulk transitions are shown in black.  The colored points indicate finite-temperature transitions determined primarily from the RG-blocked Polyakov loop and eigenvalue density.}
\end{figure}

We begin by showing in \fig{fig:phases} both finite-temperature transitions (colored points) as well as the first-order bulk transitions (black lines) that separate the \Sb phase from the weak-coupling phase.
The bulk transitions are determined from the \Sb order parameters.\cite{Cheng:2011ic}.
We also use the Dirac eigenvalue density $\rho(\la)$ to identify both bulk and finite-temperature transitions.
In addition, we discovered that by measuring observables such as the Polyakov loop on RG-blocked lattices, we can greatly enhance signals of the finite-temperature transition, without distorting the physics.\cite{Schaich:2012fr}

For $N_f = 12$ in the right panel of \fig{fig:phases}, the finite-temperature transitions occur at the same $\be_F$ as the \Sb bulk transitions.
That is, we move directly from the \Sb phase into a chirally symmetric weak-coupling phase, as expected for an IR-conformal system.

For $N_f = 8$, at sufficiently large fermion mass $m$ there is a region in between the \Sb phase and the chirally symmetric weak-coupling phase, where the lattice systems are confined and chirally broken.
For these $m$, the locations of the finite-temperature transitions change with $N_t$ as expected for QCD-like systems.
However, as we approach the chiral limit (which is required to explore IR-conformal or approximately-conformal dynamics), this scaling is lost and all the finite-temperature transitions run into the \Sb phase, with no clear sign of spontaneous chiral symmetry breaking.

At very small mass the finite spatial volume of the lattice can distort the finite-temperature phase diagram.
We are currently comparing $24^3\X12$ and $36^3\X12$ simulations to study these effects, with preliminary results indicating complete consistency for $m \geq 0.005$.
It is interesting to note that we are able to simulate directly in the chiral limit on $24^3\X48$ volumes everywhere in the weak-coupling and \Sb phases, and on $32^3\X64$ volumes for all couplings that we have tested so far ($\be_F \geq 4.8$).
This is very different from the usual QCD-like behavior, which we have checked with $N_f = 4$ simulations.
A lattice action where stronger couplings can be reached, before encountering a bulk transition, would help resolve this puzzle.

\section{\label{sec:WMCRG}The step scaling function with Wilson-flowed MCRG} 
\begin{figure}[tb]
  \centering
  \includegraphics[width=0.55\linewidth]{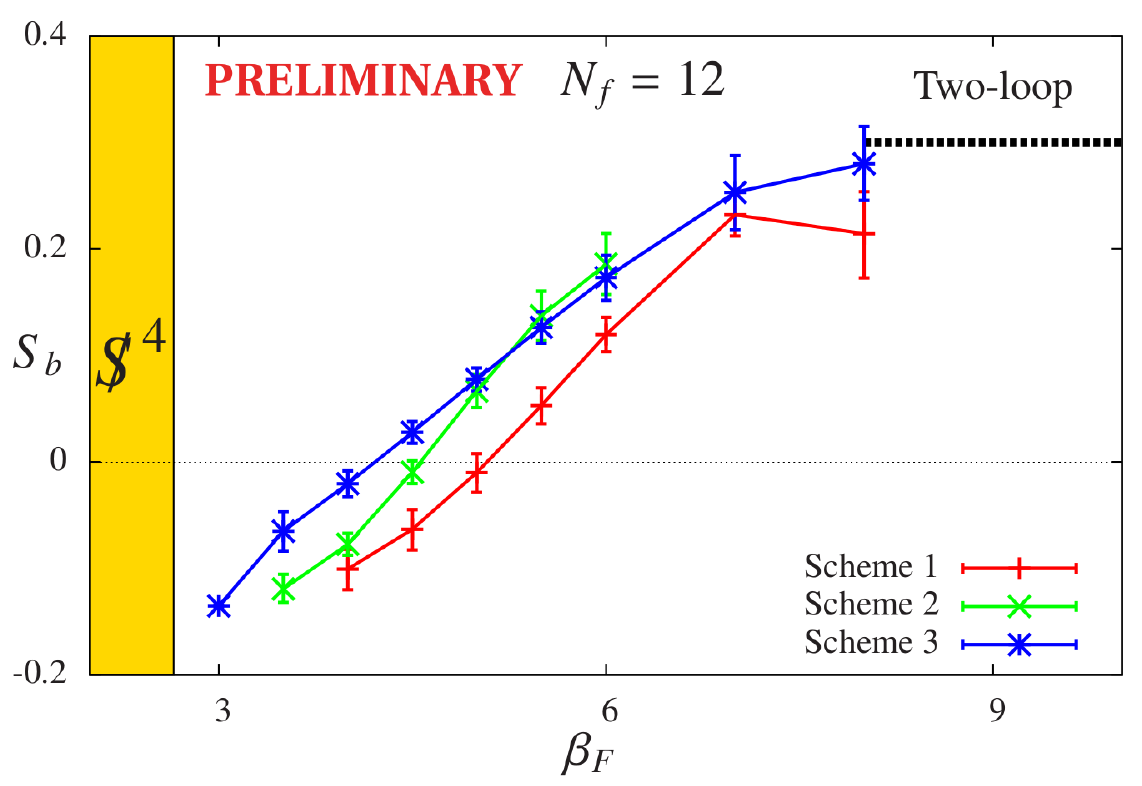}
  \caption{\label{fig:WMCRG} Preliminary results for the $N_f = 12$ bare step-scaling function $s_b$ from Wilson-flowed MCRG two-lattice matching, illustrating the existence of an IR fixed point.  We match $24^3\X48$ to $12^3\X24$ lattice volumes using three different renormalization schemes corresponding to HYP smearing parameters $\al = (0.6, 0.2, 0.2)$, $\al = (0.6, 0.3, 0.2)$ and $\al = (0.65, 0.3, 0.2)$, respectively.  The dashed line is the perturbative prediction for asymptotically weak coupling.}
\end{figure}

Running coupling studies can connect the well-understood perturbative regime to the strong-coupling region where chiral symmetry breaking or IR-conformal dynamics emerges.
We use Monte Carlo renormalization group (MCRG) methods to determine the bare step scaling function $s_b$.\cite{Hasenfratz:2011xn, Petropoulos:2012mg}
In the past we used an MCRG two-lattice matching approach that required independent optimization of the RG blocking transformation at every gauge coupling.
This predicted a step scaling function that was a composite, probing a different renormalization scheme at every coupling.

Recently we developed a new technique that combines Wilson-flow smearing with MCRG two-lattice matching.\cite{Petropoulos:2012mg}
Our new method starts with a short Wilson flow integration before the RG blocking steps.
The Wilson flow moves the configurations in the action-space without changing their infrared properties.
We use the flow time as the optimization parameter to approach the renormalized trajectory of any fixed RG blocking transformation.
This way we determine a step scaling function that corresponds to a unique RG \be function.

By changing the blocking transformation we can predict the \be function in different renormalization schemes.
\fig{fig:WMCRG} shows our preliminary results for the $N_f = 12$ system using three blocking transformations.
Each blocking transformation involves two sequential HYP smearings, with different smearing parameters specified in the caption.
All three renormalization schemes show an infrared fixed point (IRFP), and we observe the location of the IRFP to depend on the scheme.

The Wilson-flowed MCRG method is fully non-perturbative and can be applied to any other model, even at strong coupling.
There is no need for special boundary conditions, allowing configurations to be reused.
The ability to compare different renormalization schemes is also valuable, and we can gain additional numerical control by choosing a scheme that moves the IRFP to a convenient coupling.

\section{\label{sec:eig}Dirac eigenmodes and the anomalous dimension} 
\begin{figure}[tb]
  \includegraphics[width=0.45\linewidth]{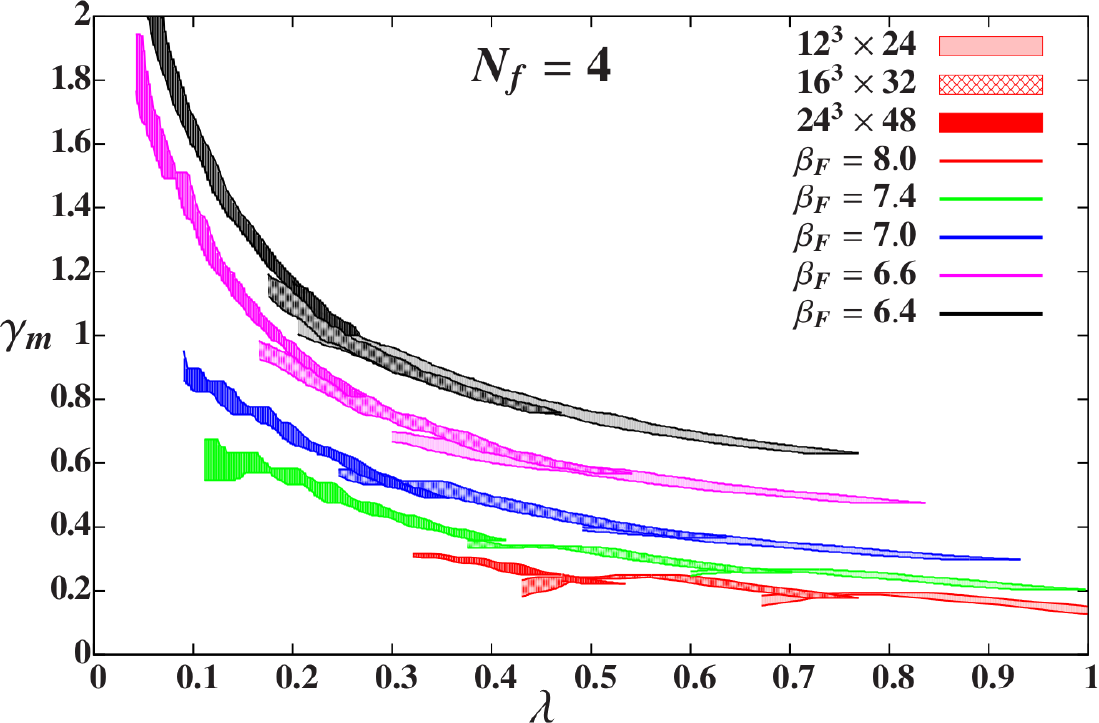}\hfill
  \includegraphics[width=0.45\linewidth]{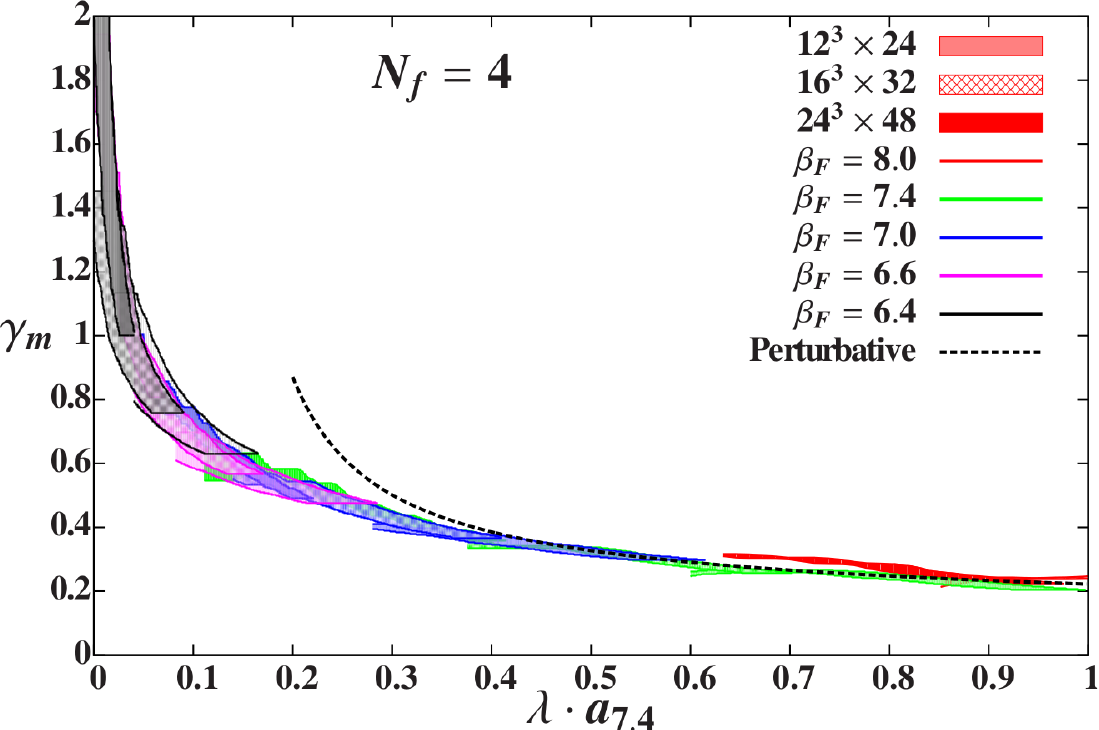}
  \caption{\label{fig:eig_4} Eigenmode scaling predictions for the scale-dependent mass anomalous dimension $\ga_m$ for $N_f = 4$.\protect{\cite{Cheng:2013eu}}  In the right panel we rescale \la to a common lattice spacing.}
\end{figure}

Our studies of the Dirac eigenvalue spectrum turned out to be particularly fruitful.
We have shown how to extract the energy-dependent mass anomalous dimension $\ga_m(\la)$ from the RG-invariant eigenmode number $\nu(\la)$,\cite{Hasenfratz:2012fp} and found that combining multiple lattice volumes and gauge couplings provides robust predictions across a wide range of energy scales.\cite{Cheng:2013eu}
We tested our procedure with the 4-flavor system, illustrated in \fig{fig:eig_4}.
The left panel shows $\ga_m$ vs.\ \la for several values of $\be_F$; in the right panel we rescale \la to a common lattice spacing $a_{7.4}$ corresponding to $\be_F = 7.4$.
We obtain a universal scaling curve that connects the one-loop asymptotic behavior at large \la to the onset of chiral symmetry breaking at small $\la$.
These results give us great confidence in our method.

\begin{figure}[tb]
  \includegraphics[width=0.45\linewidth]{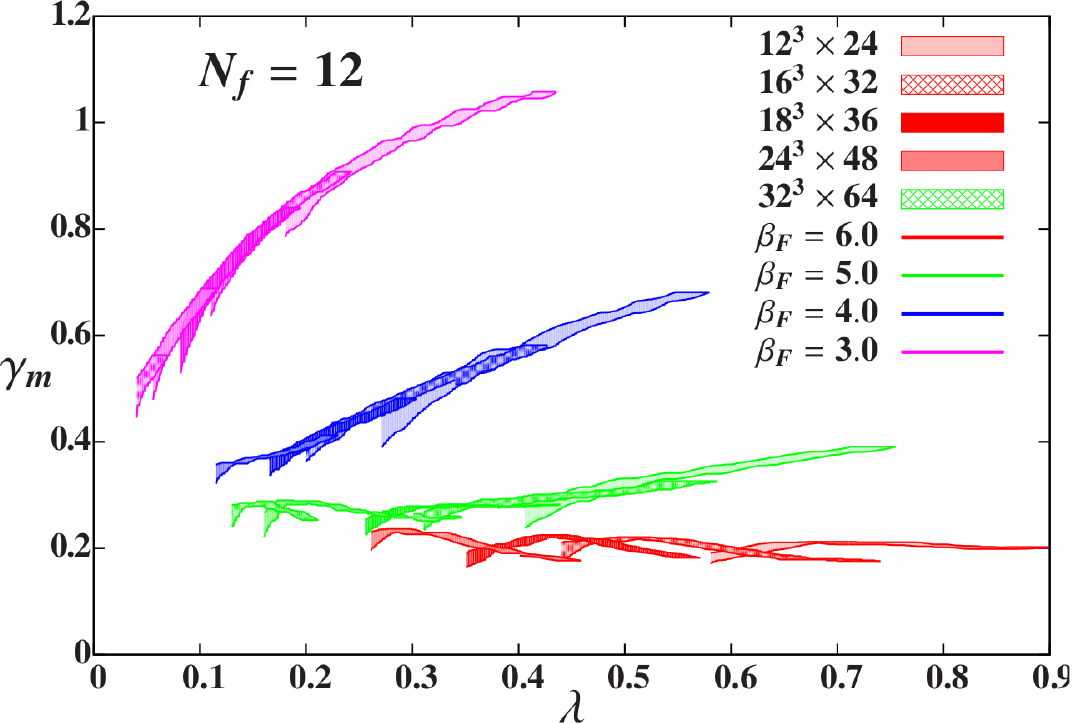}\hfill
  \includegraphics[width=0.45\linewidth]{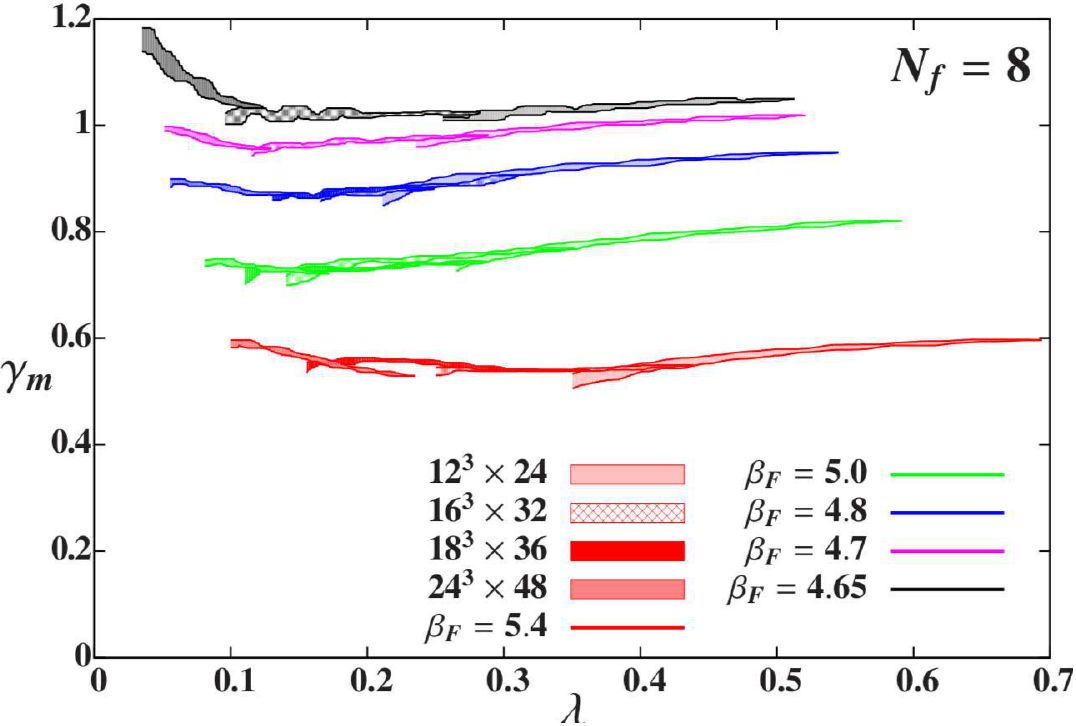}
  \caption{\label{fig:eig} Eigenmode scaling predictions for the scale-dependent mass anomalous dimension $\ga_m$ for $N_f = 12$ (left) and $N_f = 8$ (right).\protect{\cite{Cheng:2013eu}}}
\end{figure}

\fig{fig:eig} illustrates our results for the 12- and 8-flavor systems.
The behavior of $N_f = 12$ in the left panel is very different from the 4-flavor case.
At the stronger couplings $\be_F = 3.0$ and 4.0 the anomalous dimension increases with the energy scale, which can be considered a sort of backward flow.
We start seeing asymptotically-free dynamics only for $\be_F > 6.0$.
The data are consistent with an IRFP around $\be_F = 5.0$, and contradict a chirally broken scenario.
By extrapolating the predictions at different $\be_F$ to the limit $\la \to 0$ we predict a common value $\ga_m^{\star} = 0.32(3)$ that we identify as the scheme-independent mass anomalous dimension at the IR fixed point.
A prominent feature of our 12-flavor results is the strong dependence on the irrelevant, but apparently slowly-running, gauge coupling.
With $24^3\X48$ and $32^3\X64$ lattice volumes, we can only access $\la \gsim 0.1$ before encountering finite-volume effects.
The wide variation $0.2 \lsim \ga_m \lsim 0.6$ predicted by different $\be_F$ in this range of \la demonstrates the importance of considering several gauge coupling values, and may be relevant to other analyses.

The right panel of \fig{fig:eig} shows our results for $N_f = 8$.
Here the appearance of the \Sb phase around $\be_F \approx 4.6$ restricts us to a much smaller range of $\be_F$ than for $N_f = 12$.
Since none of our 8-flavor calculations show spontaneous chiral symmetry breaking (even on volumes as large as $24^3\X48$ and $32^3\X64$), we carried out this analysis with $m = 0$.
For fixed $\be_F$ we find hardly any energy dependence in our results for $\ga_m$.
At the same time, we observe significant changes as we vary the coupling.
These results suggest that the $N_f = 8$ system is very close to the edge of the conformal window.

If our $N_f = 8$ simulations probe the $\epsilon$-regime of a chirally broken system, then the distribution of the lowest eigenmodes could be compared with random matrix theory to predict the chiral condensate $\Si$.
However, our data are not consistent with $\epsilon$-regime scaling; the low-lying eigenmodes do not scale linearly with the volume, but rather with an exponent consistent with the anomalous dimension predicted by the mode number analysis.
With our present 8-flavor data we cannot distinguish walking dynamics from IR conformality.

\section{\label{sec:pbp}Chiral condensate} 
\begin{figure}[tb]
  \centering
  \includegraphics[width=0.45\linewidth]{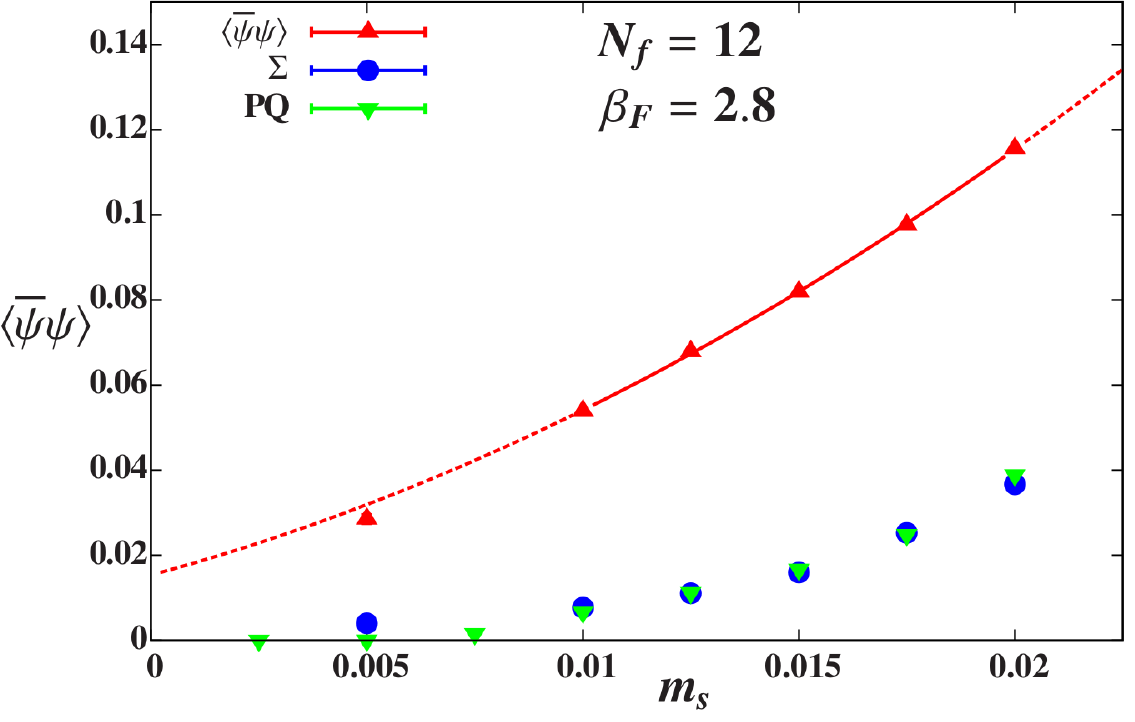}\hfill
  \includegraphics[width=0.45\linewidth]{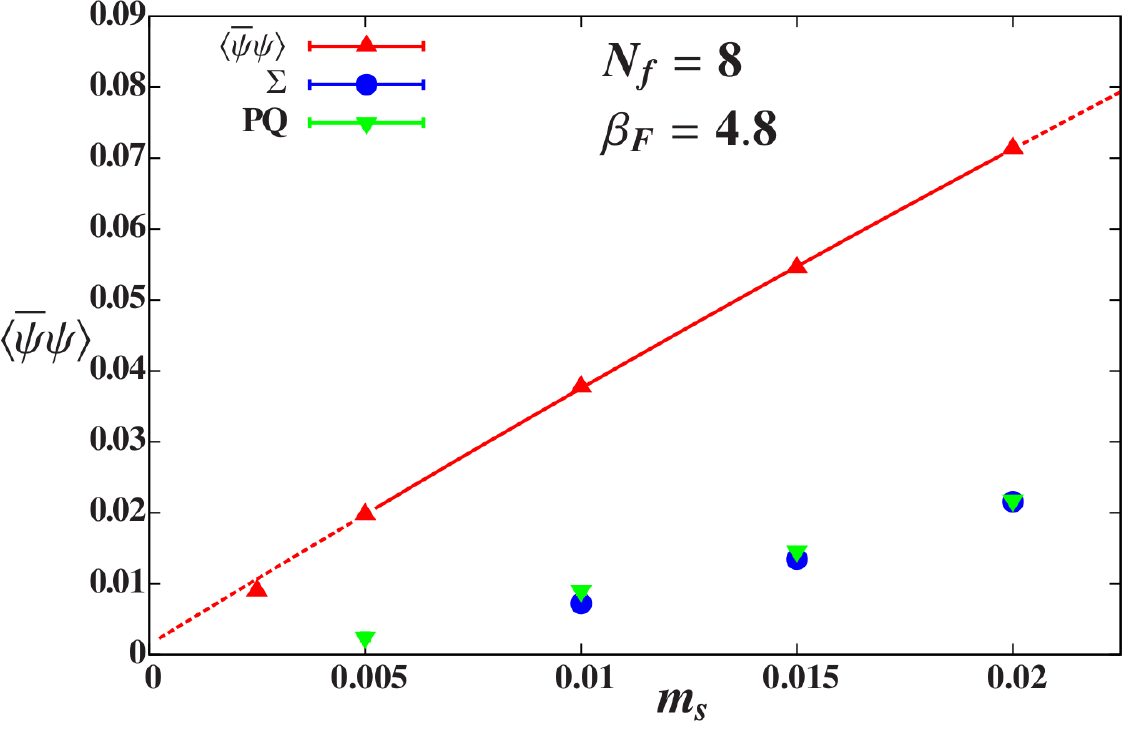}
  \caption{\label{fig:Sigma_pbp} Direct measurements of \pbp with $m_v = m_s$ (red) compared to $\Si \equiv \pi\rho(0)$ (blue) and $\Si_{PQ}$ (green).  Left: $N_f = 12$ at $\be_F = 2.8$, with the line a quadratic fit to \pbp for $0.01 \leq m_s \leq 0.02$.  Data with $m \geq 0.01$ have negligible finite-volume effects.  Right: $N_f = 8$ at $\be_F = 4.8$, with the line a quadratic fit to \pbp for $0.005 \leq m_s \leq 0.02$ where finite-volume effects are negligible.}
\end{figure}

We typically calculate \pbp with equal sea and valence fermion masses, $m = m_s = m_v$.
In the $m = 0$ chiral limit, \pbp is the order parameter of spontaneous chiral symmetry breaking.
At non-zero fermion mass, however, \pbp is dominated by a UV-divergent term $\propto m_v / a^2$ (\fig{fig:Sigma_pbp}), which complicates the $m \to 0$ chiral extrapolation.
We are investigating two alternative ways of reducing this sensitivity to the fermion mass.
The first approach is to determine the chiral condensate from the low-lying eigenvalue density of the massless Dirac operator, $\Si \equiv \pi \rho(0)$ based on the Banks--Casher relation.
Second, we carry out partially-quenched calculations with several $m_v \neq m_s$, and define $\Si_{PQ} \equiv \lim_{m_v \to 0} \pbp_{PQ}$ with fixed $m_s$.

These two approaches give consistent results that are significantly different from direct measurements of \pbp with $m_v = m_s$, as shown in \fig{fig:Sigma_pbp}.
For the 12-flavor system at the rather strong coupling $\be_F = 2.8$ (left panel), polynomial extrapolations of \pbp predict a non-zero value in the limit $m_s \to 0$, but this result is inconsistent with the data for $\Si$.
Since \Si is free of ultraviolet divergences and therefore a more reliable observable, we consider this additional evidence that the 12-flavor system is IR conformal, with vanishing chiral condensate.

Our preliminary results for $N_f = 8$ at $\be_F = 4.8$ are shown in the right panel of \fig{fig:Sigma_pbp}.
At this coupling, the chiral condensate obtained from extrapolating \pbp is very small, and could be consistent with the prediction from $\Si$.
Even so, we are cautious of such an extrapolation over at least an order of magnitude, which is caused by non-zero valence mass contributions dominating $\pbp$.

\section{Conclusion} 
These proceedings briefly overviewed our ongoing work to synthesize results from wide-ranging lattice investigations of SU(3) gauge theories with $N_f = 8$ and 12.
We presented a new Wilson-flowed MCRG method as well as a new technique to extract the mass anomalous dimension $\ga_m$ from the Dirac eigenvalue spectrum.
Both of these approaches indicate the presence of an IR fixed point for $N_f = 12$, and the Dirac spectrum predicts the scheme-independent $\ga_m^{\star} = 0.32(3)$ at the IRFP.
The 12-flavor lattice phase diagram and chiral condensate are also consistent with IR conformality.
For $N_f = 8$ our results are more puzzling and require further investigation.
The finite-temperature transitions run into the \Sb lattice phase, with no clear signs of spontaneous chiral symmetry breaking, while the Dirac spectrum predicts a large anomalous dimension across a wide range of energy scales.

\paragraph{Acknowledgments:} This research was partially supported by U.S.~DOE grants DE-FG02-04ER41290 and DE-AC05-06OR23100.
Numerical calculations were carried out on the HEP-TH and Janus clusters at the University of Colorado, the latter supported by U.S.~NSF grant CNS-0821794; at Fermilab under the auspices of USQCD supported by the DOE SciDAC program; and at the San Diego Computing Center through XSEDE supported by NSF grant OCI-1053575.

\raggedright
\bibliographystyle{ws-procs975x65}
\bibliography{hasenfratz}
\end{document}